\documentclass{article}
\usepackage{spconf,amsmath,graphicx,hyperref}
\usepackage{multirow}
\usepackage{booktabs}
\usepackage{xcolor}

\title{Addressing Gradient Misalignment in Data-Augmented Training for Robust Speech Deepfake Detection}
\name{Duc-Tuan Truong$^1$, Tianchi Liu$^{2}$, Junjie Li$^{3}$, Ruijie Tao$^{2, \dag}$\thanks{$\dag$ Co-corresponding author.}, Kong Aik Lee$^{3,\dag}$, Eng Siong Chng$^1$}
\address{$^1$Nanyang Technological University, Singapore\quad$^2$National University of Singapore, Singapore \\ $^3$The Hong Kong Polytechnic University, Hong Kong \\}
\begin{document}
\ninept
\maketitle
\begin{abstract}
In speech deepfake detection (SDD), data augmentation (DA) is commonly used to improve model generalization across varied speech conditions and spoofing attacks. However, during training, the backpropagated gradients from original and augmented inputs may misalign, which can result in conflicting parameter updates. These conflicts could hinder convergence and push the model toward suboptimal solutions, thereby reducing the benefits of DA. To investigate and address this issue, we design a dual-path data-augmented (DPDA) training framework with gradient alignment for SDD. In our framework, each training utterance is processed through two input paths: one using the original speech and the other with its augmented version. This design allows us to compare and align their backpropagated gradient directions to reduce optimization conflicts. Our analysis shows that approximately 25\% of training iterations exhibit gradient conflicts between the original inputs and their augmented counterparts when using RawBoost augmentation. By resolving these conflicts with gradient alignment, our method accelerates convergence by reducing the number of training epochs and achieves up to an 18.69\% relative reduction in Equal Error Rate on the In-the-Wild dataset compared to the baseline\footnote{Code and models are available at \href{https://github.com/ductuantruong/dpda_ga}{github.com/ductuantruong/dpda\_ga}}.
\end{abstract}
\section{Introduction}
\label{sec:intro}
Speech deepfake detection (SDD) models perform well under seen conditions but struggle to generalize to unseen scenarios. They often suffer from significant accuracy drops under new attack types, acoustic conditions, or noise that can hide synthetic speech artifacts \cite{kwok25_interspeech, 10832142,liu24m_interspeech,qq}. To address this challenge, data augmentation (DA) is a straightforward strategy to increase training variability for SDD. Common DA techniques in speech processing include room impulse response (RIR) \cite{rirs,rir_aug}, environmental noise addition \cite{musan}, and masking \cite{specaug,mask_aug}. Based on that, RawBoost \cite{rawboost} is designed for the SDD task by applying signal processing to distort raw waveforms, which enhances robustness to acoustic variations. Recent studies have explored different SDD augmentation strategies. For example, \cite{astrid2024targeted} proposes an adversarial attack-inspired method that produces pseudo-fake samples near decision boundaries. CpAug \cite{cpaug} applies a copy–paste strategy to enhance intra-class diversity. Additionally, \cite{huang2025generalizable} explores embedding-level augmentation to diversify the feature space and promote more generalizable representations.



While DA improves the robustness of SDD, prior work \cite{rir_aug} has shown that SDD models trained with DA still suffer from performance degradation when evaluated under new acoustic conditions within the same test set. This suggests that DA alone is insufficient and better results may be achieved by helping the model focus on distinguishing bona fide from spoofed audio rather than overfitting to acoustic variations introduced by augmentation. We hypothesize that the sensitivity of SDD models to DA perturbations leads to different loss surfaces between original and augmented inputs. As a result, gradients backpropagated from these two input types may point in different directions, causing gradient conflicts during joint updates. These conflicts can slow down convergence, and push optimization toward suboptimal solutions. Similar issues have been observed in multi-task learning \cite{gradconflict}, where gradients from different tasks can be conflicting and harm training. Gradient alignment methods were originally introduced in multi-task training to reduce such conflicts and stabilize optimization \cite{pcgrad,gradvac,cagrad}. These methods were later extended to domain generalization to enforce gradient consistency across domains \cite{dg_ga}.

Motivated by these observations, we propose a dual-path data-augmented (DPDA) training framework with gradient alignment for speech deepfake detection (SDD). In this framework, the model is trained using both original and augmented version of each training utterance, and a gradient alignment method is applied when their backpropagated gradients conflict. Although gradient alignment methods have been studied in other contexts such as multi-task learning and domain generalization, our approach is distinct in that it aligns gradients between two inputs derived from the same utterance but modified through DA. This alignment encourages consistent parameter updates and helps the model focus on spoof-related cues rather than overfitting to augmentation artifacts. To the best of our knowledge, this is the first work to investigate gradient conflicts during data-augmented training in the context of SDD. Our study provides the following key findings: (i) gradient conflicts between the original and augmented inputs are linked to differences in their loss landscapes; (ii) such conflicts occur frequently during DPDA training; and (iii) resolving them consistently improves performance across multiple SDD architectures, augmentation strategies, and benchmark datasets over both our DPDA baseline and conventional single-path data-augmented training. Finally, our approach is architecture-agnostic and can be easily integrated with existing models and signal-level data augmentation techniques.

\label{sec:method}
\begin{figure*}[t!]
  \centering
  \includegraphics[width=\textwidth]{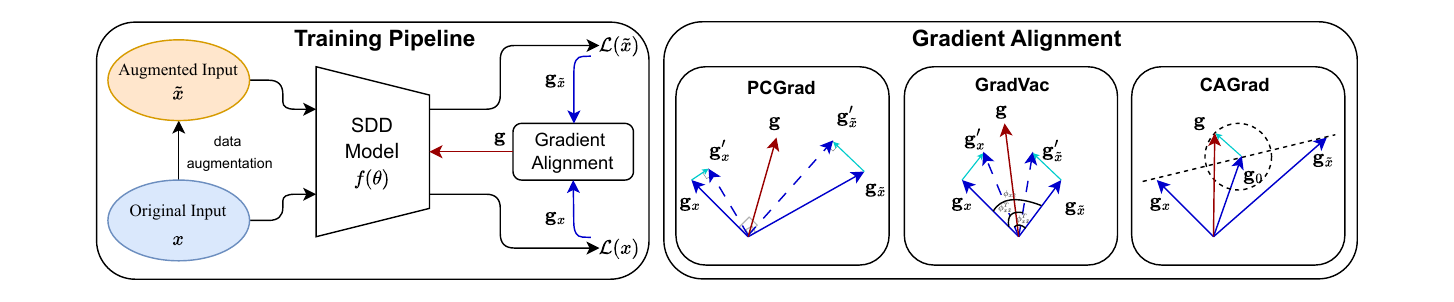}
  \vspace{-9mm}
  \caption{Overview of the dual-path data-augmented (DPDA) training framework with gradient alignment for SDD. The left panel illustrates the overall training framework, wherein both original and augmented inputs are processed in parallel. The right panel compares three gradient alignment methods, including PCGrad \cite{pcgrad}, GradVac \cite{gradvac}, and CAGrad \cite{cagrad}, which are used to resolve gradient conflicts.}
  \label{fig:main_arch}
\end{figure*}
\section{Methodology}
\begin{figure}[t!]
  \centering
  \vspace{-7mm}
  \includegraphics[width=0.95\linewidth]{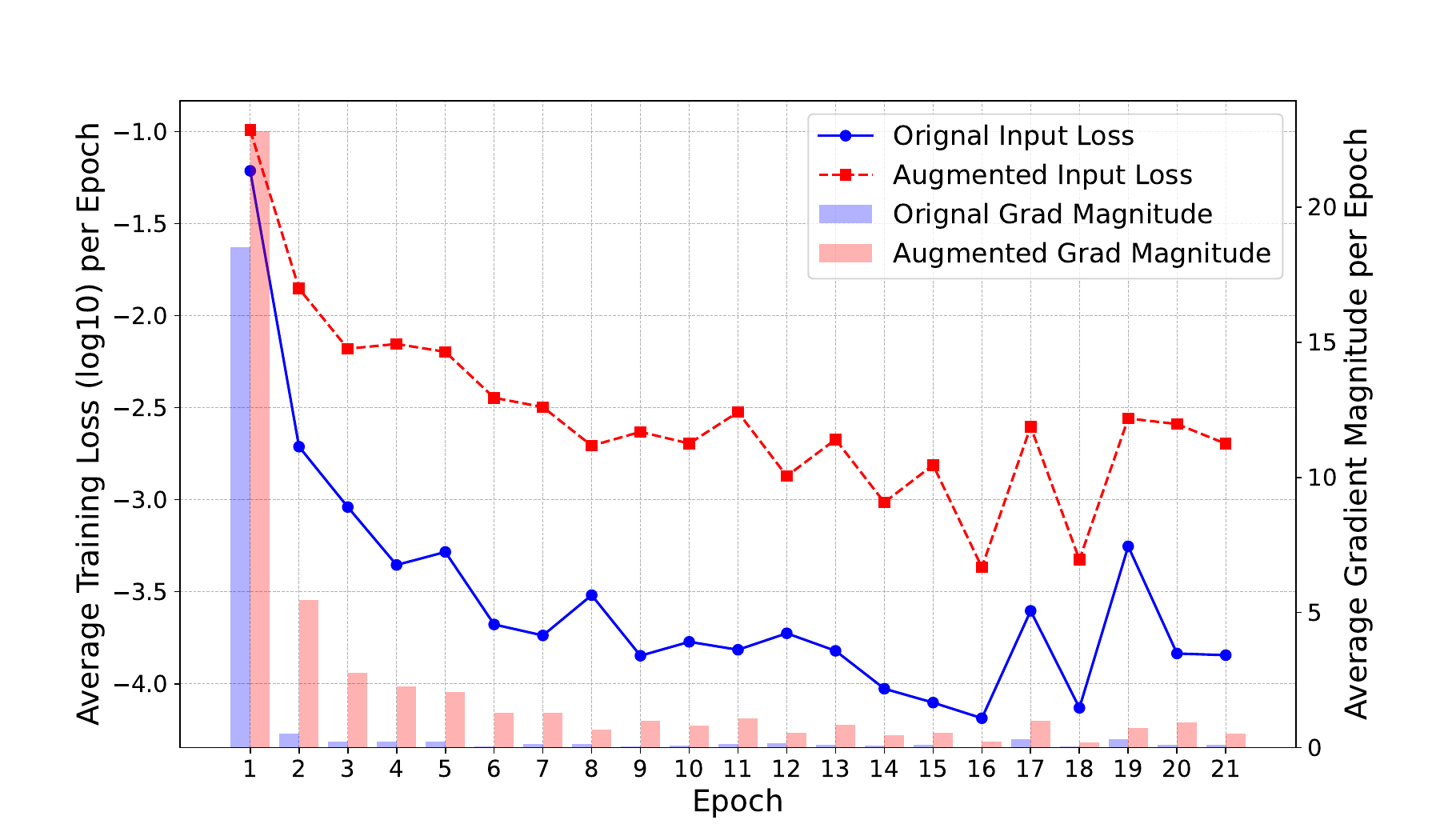}
  \vspace{-6mm}
  \caption{Training loss and backpropagated gradient norm of orignal $x$ and augmented $\tilde{x}$ inputs during the DPDA training of XLSR-Conformer-TCM model.}
  \label{fig:loss_grad_plot}
\end{figure}
\subsection{The proposed training framework}
Figure \ref{fig:main_arch} illustrates our proposed DPDA training framework with gradient alignment. Let $x$ denote the original input sample and the augmented counterpart $\tilde{x}$ generated through an input-level data augmentation method. Both inputs are formed as a mini-batch and passed through a shared SDD model $f(\theta)$ with parameters $\theta$ to compute the loss $\mathcal{L}$. In conventional training, the gradients from both inputs are aggregated directly as $\mathbf{g}=\nabla_{\theta}\mathcal{L}([x,\tilde{x}])$. However, although $x$ and $\tilde{x}$ come from the same utterance, their gradients may differ, and direct aggregation can introduce conflicts that hinder optimization. 

To investigate this, we compare the training losses $\mathcal{L}(x)$ and $\mathcal{L}(\tilde{x})$ alongside the gradient magnitudes computed from each input path. As shown in Figure \ref{fig:loss_grad_plot}, there is a clear pattern that the training loss of the augmented input path is higher than the original one, hence leads to a larger gradient norms of the augmented input compared to the original one. This imbalance can cause the model updates to be dominated by the augmented input. As a result, this dominance may bias the model toward learning augmentation-specific patterns, rather than focusing on spoof-related cues. The issue can become worse when the gradients from $x$ and $\tilde{x}$ point in different directions, which can introduce conflicts during optimization. 

To further investigate the source of gradient conflicts between original and augmented inputs, we visualize the local loss surfaces of the model for both inputs at an intermediate point during training. Following the method introduced by \cite{visualloss}, we approximate the loss landscape around the model parameters by perturbing them along two randomly sampled, orthogonal directions in parameter space. The directions are normalized and scaled, and the loss is evaluated on a grid around the current parameter point, forming a 2D surface plot. This approach provides an interpretable visualization of the local geometry of the loss landscape.

\begin{figure}[t!]
  \centering
  \vspace{-2mm}
  \includegraphics[width=\linewidth]{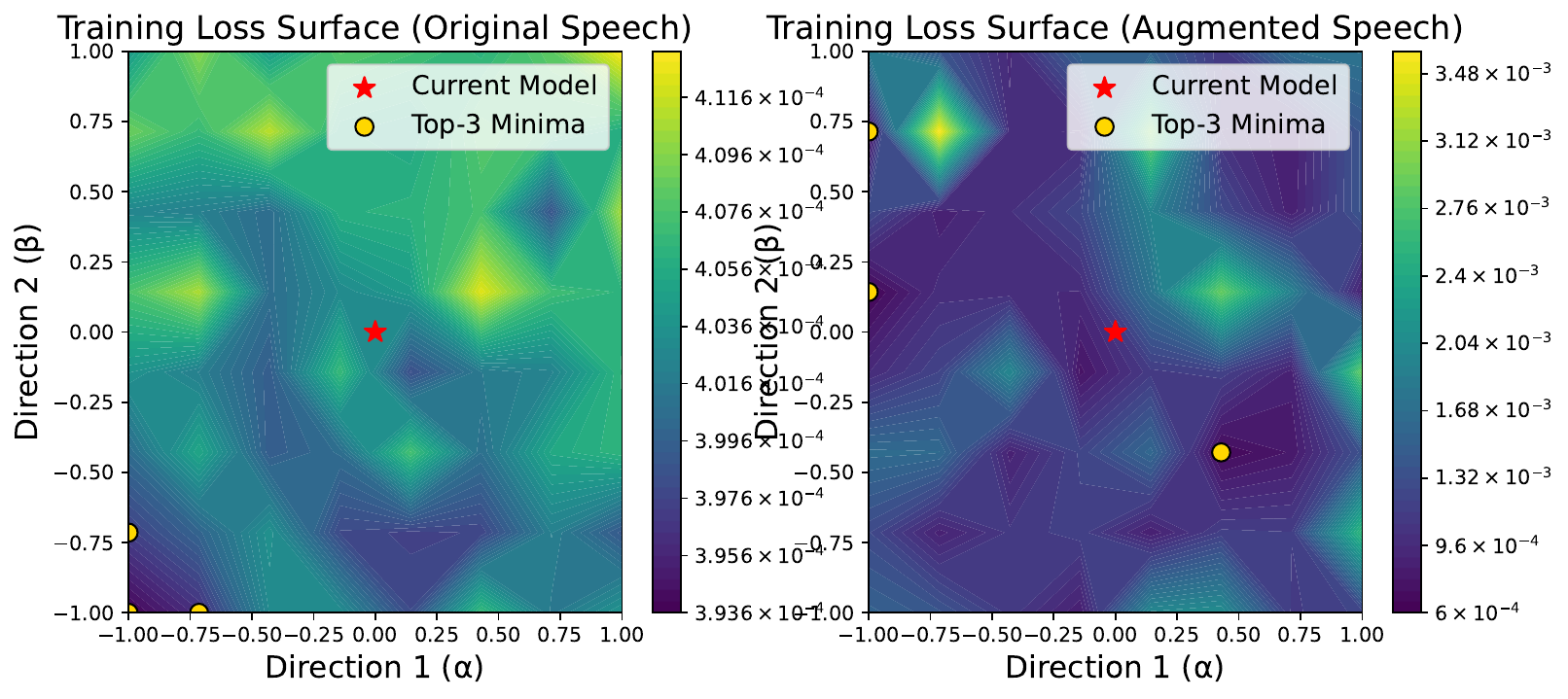}
  \vspace{-8mm}
  \caption{Loss surface visualization for original and augmented inputs of the XLSR-Conformer-TCM model during mid-training}
  \label{fig:loss_surface}
\end{figure}

Figure \ref{fig:loss_surface} shows loss surfaces for both the original and augmented input paths. The loss surface of the original input is relatively smooth, with fewer nearby suboptimal points that have lower training loss than the current model parameters. In contrast, the augmented input produces a more complex surface, with several sharp valleys and suboptimal points, suggesting a more complicated optimization region. We identify the top-3 lowest loss points on each surface to approximate the directions in which the model parameters would ideally move to reduce the training loss, as guided by gradient descent. As can be seen, the directions toward corresponding minima points on the original and augmented loss surfaces are not well aligned, indicating a mismatch in their optimization trajectories. These differences in surface geometry and descent direction reinforce our hypothesis that gradient conflicts arise due to the distinct loss landscapes induced by data augmentation. These findings highlight the need for gradient alignment during training to keep the optimization direction from both inputs consistent and help the model focus on detecting spoofed speech rather than being influenced by input variations.

\subsection{Gradient alignment methods}
Let $\mathbf{g}_x = \nabla_{\theta}\mathcal{L}(x)$ and $\mathbf{g}_{\tilde{x}} = \nabla_{\theta}\mathcal{L}(\tilde{x})$ denote the gradients obtained from the original and augmented versions of the same utterance, respectively. A general gradient alignment framework consists of two steps: (i) defining a criterion to determine whether $\mathbf{g}_x$ and $\mathbf{g}_{\tilde{x}}$ are in conflict, and (ii) applying a mechanism to adjust the gradients when a conflict is detected. The aligned gradients, denoted as $\mathbf{g}^\prime_x$ and $\mathbf{g}^\prime_{\tilde{x}}$, are then combined to form the final update direction $\mathbf{g}$. In this work, we examine three gradient alignment methods, including PCGrad \cite{pcgrad}, GradVac \cite{gradvac}, and CAGrad \cite{cagrad}, which differ in how gradient conflicts are detected and subsequently adjusted.

\subsubsection{PCGrad \cite{pcgrad}}
In PCGrad, two gradients are considered conflicting if their inner product is negative (i.e. $\left\langle\mathbf{g}_x, \mathbf{g}_{\tilde{x}}\right\rangle<0$), indicating opposing update directions. To resolve this, each gradient is projected onto the normal plane of the other to remove the conflicting component, as defined:
\begin{equation}
    \mathbf{g^\prime}_x = \mathbf{g}_x-\frac{\left\langle\mathbf{g}_x, \mathbf{g}_{\tilde{x}}\right\rangle}{\left\|\mathbf{g}_{\tilde{x}}\right\|^2} \mathbf{g}_{\tilde{x}}
\end{equation}
A similar operation is applied to $\mathbf{g}_{\tilde{x}}$. This alignment prevents the gradients from moving in opposing directions, allowing stable updates across both original and augmented inputs. As a result, the SDD training can reduce the gradient disturbance introduced by DA and focus more on the authenticity of the speech utterance.
\subsubsection{GradVac \cite{gradvac}}
GradVac extends gradient alignment by adaptively setting target cosine similarity goals between gradient pairs based on their expected relationship. Unlike PCGrad, which only acts when gradients conflict, GradVac aligns gradients whenever their cosine similarity $\phi_{x \tilde{x}} = \frac{\left\langle\mathbf{g}_x, \mathbf{g}_{\tilde{x}}\right\rangle}{|\mathbf{g}_x||\mathbf{g}_{\tilde{x}}|}$ falls below a positive target $\phi^T_{x\tilde{x}}$. When needed, GradVac modifies $\mathbf{g}_x$ via a linear combination of both gradients to achieve the target similarity:
\begin{equation}
\small \mathbf{g^\prime}_x=\mathbf{g}_x+\frac{\left\|\mathbf{g}_x\right\|\left(\phi_{x \tilde{x}}^T \sqrt{1-\phi_{x \tilde{x}}^2}-\phi_{x \tilde{x}} \sqrt{1-\left(\phi_{x \tilde{x}}^T\right)^2}\right)}{\left\|\mathbf{g}_{\tilde{x}}\right\| \sqrt{1-\left(\phi_{x \tilde{x}}^T\right)^2}} \cdot \mathbf{g}_{\tilde{x}}
\end{equation}
The target similarity $\phi^T_{x\tilde{x}}$ is updated as an exponential moving average of the positive similarities $\phi_{x \tilde{x}}$ observed during training. Therefore, GradVac not only avoids conflicts but also encourages closer alignment between gradients from the dual-path inputs.

\subsubsection{CAGrad \cite{cagrad}}
Conflict-Averse Gradient Descent (CAGrad) offers a principled solution to align gradients from the original input and its augmented counterpart by finding an update that balances improving both objectives while minimizing conflicts. Unlike PCGrad and GradVac, which remove the gradient conflict by projection, CAGrad seeks the final gradient $\mathbf{g}$ close to the conventional update gradient $\mathbf{g}_0=\nabla_{\theta}\mathcal{L}([x,\tilde{x}])$ while maximizing the minimum improvement over both gradients by solving the following:
\begin{equation}
\max_{\mathbf{g}} \min \left\{\left\langle\mathbf{g}_x, \mathbf{g}\right\rangle,\left\langle\mathbf{g}_{\tilde{x}}, \mathbf{g}\right\rangle\right\} \; \text{s.t.} \; \left\|\mathbf{g}-\mathbf{g}_0\right\| \leq c\left\|\mathbf{g}_0\right\|
\label{eq:cagrad}
\end{equation}
where $c \in[0,1)$ is a hyperparameter controlling the proximity to $\mathbf{g}_0$. By solving this convex optimization problem, CAGrad aims to find an optimal updated gradient between the gradients of the original and augmented input.

\section{EXPERIMENTAL SETUP}
\label{sec:experiment}
\subsection{Dataset and metrics}
We train and validate our method on the ASVspoof2019 Logical Access (LA) dataset \cite{asvspoof2019}, which consists of bona fide and spoofed speech generated by diverse text-to-speech (TTS) and voice conversion (VC) systems. For evaluation, three datasets including ASVspoof2021 DF (21DF) \cite{asvspoof2021}, In-the-Wild (ITW) \cite{itw} and the Fake-or-Real (FoR) \textit{norm-test} subset\footnote{\href{https://www.kaggle.com/datasets/mohammedabdeldayem/the-fake-or-real-dataset}{kaggle.com/datasets/the-fake-or-real-dataset}} are utilized. The 2021DF dataset includes a wide range of spoofing attacks and simulates real-world conditions including codec compression and channel variability. The ITW and FoR datasets further test robustness using real-world audio sourced from online media and podcasts. We report the standard Equal Error Rate (EER) as the main metric.
\subsection{Implementation details}
We evaluate the proposed method on multiple model architectures, including XLSR-AASIST \cite{xlsr_aasist}, XLSR-TCM-Conformer \cite{conformer_tcm}, and XLSR-Mamba \cite{xlsr_mamba}, to validate its generalization. All experiments follow the original training settings of each model. However, the dual-path data-augmented training receives both original and augmented inputs, which roughly doubling memory usage, as gradients for both need to be stored. To fit the GPU memory during training, we need to reduce the training mini-batch size from 20 (as used in the original setting) to 10, which includes 5 original and 5 augmented samples. For data augmentation, we use the widely adopted RawBoost method \cite{rawboost}. Unlike other works \cite{xlsr_aasist, conformer_tcm, xlsr_mamba, kan,kwok2025robust} that train separate models with RawBoost configuration 3 and 5 to evaluate on different test sets, we unify all experiments by applying RawBoost configuration 4. This choice saves computational resources while covering all noise types included in both configurations 3 and 5. No augmentation is applied during validation.

\section{RESULTS AND ANALYSIS}
\label{sec:result}

\subsection{Comparison of different gradient alignment methods}
When testing on the XLSR-Conformer-TCM model, Table \ref{tab:diff_ga} shows that all three examined gradient alignment methods outperform the baseline system without gradient alignment on the 21DF, ITW and FoR datasets. This highlights the benefit of using gradient alignment in data-augmented training for robust SDD since the 21DF, ITW and FoR datasets are challenging, featuring a wider variety of attack types and acoustic conditions. Notably, despite being the simplest of the tested methods originally proposed for multi-task training, PCGrad achieves the lowest EER on both 21DF and ITW. However, the reason for its superior performance remains unclear, pointing to the need for further investigation into how gradient conflicts should be resolved in context of data-augmented training. Since PCGrad achieves the best performance among the gradient alignment methods, we adopt PCGrad as the primary gradient alignment strategy for all subsequent experiments.

\begin{table}[t]
\centering
\caption{Performance comparison of XLSR-Conformer-TCM between different gradient alignment methods and the dual-path data-augmented training baseline.}
\vspace{2mm}

\begin{tabular}{lccc}
\hline

\multirow{2}{*}{\textbf{System}}                        & \multicolumn{3}{c}{\textbf{EER (\%)}}         \\ \cline{2-4} 
                            & \textbf{21DF}          & \textbf{ITW}           & \textbf{FoR} \\ \hline    
DPDA training                             & 2.11      & 7.97        & 5.31      \\ \hline
\enspace + PCGrad \cite{pcgrad}           & \textbf{1.81}      & \textbf{6.48}        & 4.47      \\
\enspace + GradVac \cite{gradvac}         & 1.83      & 7.09        & 4.81      \\
\enspace + CAGrad \cite{cagrad}           & 1.92      & 7.45        & \textbf{4.23}      \\ \hline
\end{tabular}
\label{tab:diff_ga}
\end{table}

\subsection{Experiments with different model architectures}
Table \ref{tab:diff_arch} shows that our proposed gradient alignment framework improves the performance across most test sets of the three model architectures, XLSR-AASIST, XLSR-Conformer-TCM and XLSR-Mamba. While training with the DPDA baseline led to higher EER compared to each model's original single-path results in some of evaluation results, the addition of gradient alignment not only overcomes this drop but also consistently outperforms both the reported single-path and the dual-input data-augmented training baselines. This demonstrates that our strategy can be generalized across different model architectures for speech deepfake detection. Notably, XLSR-AASIST with gradient alignment achieves the best EER on ITW (5.42\%) and FoR (3.04\%), while XLSR-Mamba with gradient alignment provides the lowest EER on 21DF (1.74\%).
\begin{table}[]
\centering
\caption{Performance comparison of our DPDA training with the gradient alignment method PCGrad across different model architectures, alongside the DPDA training baseline and other state-of-the-art models in the first block (* denotes results reported by \cite{arena}).}
\vspace{2mm}
\begin{tabular}{lccc}
\hline

\multirow{2}{*}{\textbf{System}}                        & \multicolumn{3}{c}{\textbf{EER (\%)}}         \\ \cline{2-4} 
                            & \textbf{21DF}          & \textbf{ITW}           & \textbf{FoR} \\ \hline
 XLSR-MoE \cite{xlsr_moe}                                    & 2.54                 & 9.17              & -        \\
 XLSR-SLS \cite{sls}                                         & 1.92                 & 7.46              & 5.07*        \\
 XLSR-Nes2Net-X \cite{nes2net}                               & 1.78                 & 6.60              & 6.31*        \\ 
 XLSR-AASIST-SAM \cite{sam}                                  & 3.44                 & 6.34              & 5.18        \\
 Wav2DF-TSL \cite{ustc}                                      & 1.95                 & 6.83              & -        \\
 \hline
 XLSR-AASIST \cite{xlsr_aasist}                              & 3.69                 & 10.46             & 7.46*       \\

 w/ DPDA training                                            & \textbf{1.87}        & 6.20              & 4.60            \\
                                     \enspace + PCGrad       & 2.13                 & \textbf{5.42}     & \textbf{3.04}            \\ \hline
 XLSR-Conformer-TCM \cite{conformer_tcm}                     & 2.06                 & 7.79              & 10.68*            \\                          
 w/ DPDA training                                            & 2.11                 & 7.97              & 5.31            \\
                                     \enspace + PCGrad       & \textbf{1.81}        & \textbf{6.48}     & \textbf{4.47}            \\ \hline
 XLSR-Mamba \cite{xlsr_mamba}                                & 1.88                 & 6.70              & 6.71*            \\
 w/ DPDA training                                            & 2.31                 & 7.62              & 5.39            \\
                                     \enspace + PCGrad       & \textbf{1.74}        & \textbf{6.43}     & \textbf{4.86}            \\ 
                                     \bottomrule
\end{tabular}

\label{tab:diff_arch}
\end{table}

\begin{figure}[t]
  \centering
  \includegraphics[width=0.93\linewidth]{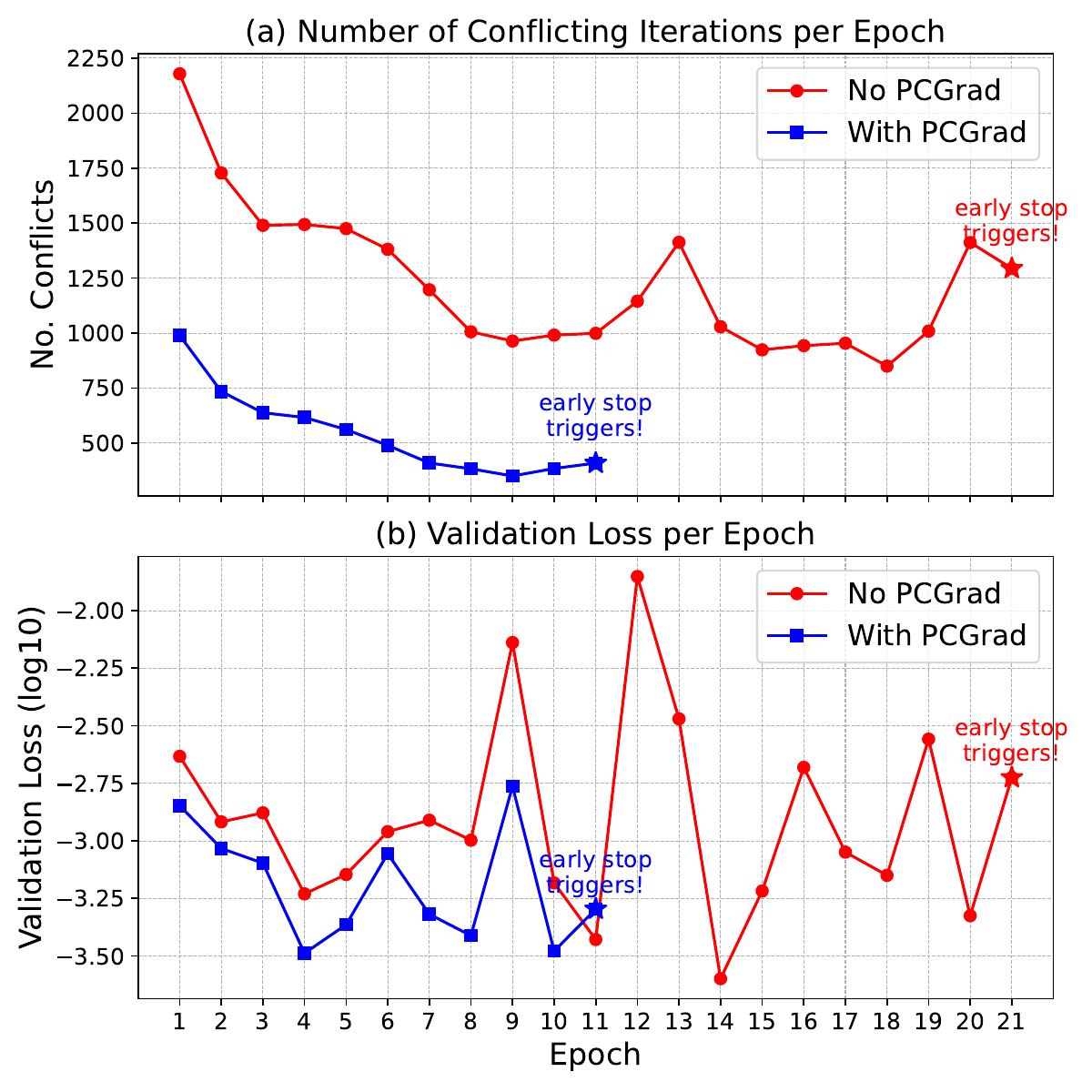}
 \vspace{-5mm}

\caption{Number of gradient conflicts and validation loss during DPDA training of the XLSR-Conformer-TCM model, with and without the PCGrad gradient alignment method. Each epoch contains 5,076 iterations.}
\label{fig:conflict_log}
\end{figure}

\begin{table}[t]
\centering
\caption{Performance of the XLSR-Conformer-TCM model comparing dual-path data-augmented training with and without gradient alignment across various data augmentation methods.}
\vspace{2mm}
\begin{tabular}{llccc}
\hline
\multirow{2}{*}{\textbf{DA type}} & \multirow{2}{*}{\textbf{System}}                        & \multicolumn{3}{c}{\textbf{EER (\%)}}         \\ \cline{3-5} 
                   &         & \textbf{21DF}          & \textbf{ITW}           & \textbf{FoR} \\ \hline

\multirow{2}{*}{RawBoost}                                                          & DPDA training       & 2.11          & 7.97           & 5.31    \\
                                                                                   & \enspace + PCGrad   & \textbf{1.81} & \textbf{6.48}  & \textbf{4.47}    \\ \hline

\multirow{2}{*}{MUSAN \& RIR}                                                      & DPDA training       & 5.45          & 23.04          & 12.02    \\
                                                                                   & \enspace + PCGrad   & \textbf{3.81} & \textbf{19.43} & \textbf{8.05}    \\ \hline
\multirow{2}{*}{\begin{tabular}[c]{@{}l@{}}MUSAN \& RIR\\ + RawBoost\end{tabular}} & DPDA training       & 1.78          & 8.10           & \textbf{2.83}    \\
                                                                                   & \enspace + PCGrad   & \textbf{1.63} & \textbf{7.19}  & 2.91    \\ \hline

\end{tabular}
\label{tab:aug_abl}
\end{table}

\subsection{Gradient conflict analysis of the dual-path data-augmented training}
In the XLSR-Conformer-TCM training experiment, Figure~\ref{fig:conflict_log} shows that the baseline dual-path data-augmented training without PCGrad suffers from nearly twice as many gradient conflict iterations as when using the gradient alignment PCGrad method. PCGrad also steadily reduces conflicts throughout training. On average, about 25\% of the baseline's training iterations suffer gradient conflict, highlighting the significance of this issue when training with data augmentation. Similarly, DPDA training with PCGrad shows a more stable and faster reduction in validation loss. This suggests that gradient alignment between original and augmented inputs improves training stability. Although gradient alignment doubles the computation per iteration by requiring gradient calculations for both original and augmented inputs, it enables the model to converge faster by 43\%. Specifically, the PCGrad-based model reaches its lowest validation loss at epoch 4, whereas the baseline reaches its minimum at epoch 14. In both experiments, training stops after seven consecutive epochs without improvement in validation loss. Overall, gradient alignment not only reduces the frequency of conflicts but also leads to faster overall convergence despite the added per-iteration cost.

\subsection{Experiments with different data augmentation methods}
Table \ref{tab:aug_abl} demonstrates that our gradient alignment pipeline consistently improves performance of XLSR-Conformer-TCM model across the three types of noise conditions: RawBoost, MUSAN Noise \cite{musan} and RIR \cite{rirs}, and their combination. In most test sets, adding PCGrad lowers the EER compared to each baseline, showing that our method can generalize well across different noise augmentation strategies. Notably, models trained with RawBoost both with and without PCGrad achieve better results than those augmented with MUSAN Noise and RIR, indicating the effectiveness of RawBoost for this task. Finally, these findings highlight that our proposed approach is robust and effective for enhancing deepfake detection under various data augmentation methods.

\section{CONCLUSION}
\label{sec:conclusion}
This paper presents a dual-path data-augmented training framework with gradient alignment for speech deepfake detection. Our analysis shows that even within the same utterance, gradients from the original and augmented inputs can conflict. Applying gradient alignment during training effectively resolves these conflicts, resulting in faster convergence and improved robustness. To demonstrate the general effectiveness, we validate our proposed method across multiple gradient alignment techniques, model architectures, and augmentation methods. Overall, this work provides empirical evidence supporting the importance of gradient alignment when training speech deepfake detection models with augmented data. It also points to future research directions, including theoretically understanding the conditions under which gradient conflicts arise and developing alignment methods tailored to data-augmented SDD training.
\vfill\pagebreak

\begingroup
\footnotesize
\bibliographystyle{IEEEbib}
\bibliography{strings,refs}
\endgroup
\end{document}